# Low-energy electron holographic imaging of gold nanorods supported by ultraclean graphene

*Jean-Nicolas Longchamp\*, Conrad Escher, Tatiana Latychevskaia & Hans-Werner Fink*

*Physics Institute, University of Zurich, Winterthurerstrasse 190, 8057 Zurich, Switzerland*

**\*Corresponding Author**

E-mail: longchamp@physik.uzh.ch

**Abstract**

An ideal support for electron microscopy shall be as thin as possible and interact as little as possible with the primary electrons. Since graphene is atomically thin and made up of carbon atoms arranged in a honeycomb lattice, the potential to use graphene as substrate in electron microscopy is enormous. Until now graphene has hardly ever been used for this purpose because the cleanliness of freestanding graphene before or after deposition of the objects of interest was insufficient. We demonstrate here by means of low-energy electron holographic imaging that freestanding graphene prepared with the Platinum-metal catalysis method remains ultraclean even after re-exposure to ambient conditions and deposition of gold nanorods from the liquid phase. In the holographic reconstruction of the gold particles the organic shell surrounding the objects is apparent while it is not detectable in SEM images of the very same sample, demonstrating the tremendous potential of low-energy electron holography for the imaging of graphene-supported single biomolecules.

**Introduction**

To image an object by means of electron microscopy, it is normally placed onto a substrate. The signal from the object support, arising from the scattering of the impinging primary electrons in transmission electron microscopy, or from the creation of secondary electrons in a scanning electron microscope, is spurious and efforts to reduce these signals have been accomplished since the development and implementation of the first electron microscopes. Ideally, for maximal contrast and resolution, one would like to have the thinnest substrate possible, made up of a low-atomic-number material, in order to reduce the interaction volume and the scattering cross-section of the incoming electrons [1,2]. The idea of using freestanding single-layer graphene as such ultimate microscopic sample carrier in electron microscopy [3–10] has been around since the isolation of single-layer graphene was achieved in 2004 by Geim and Novoselov [11,12].

Significant efforts have been undertaken in the past few years to develop techniques for preparing either exfoliated or CVD grown graphene in a freestanding form [9,13–16]. Unfortunately, the cleanliness of the prepared graphene sheets has never been satisfactory with regards to their use as sample carrier [9,10,15,17]. Only recently, it has become possible to prepare ultraclean freestanding graphene by platinum-metal catalysis [18]. Compared to previous methods, the one applied here leads to large regions, extending up to several square microns of atomically clean freestanding graphene suitable for use in electron microscopy [18–20].

Here, we show that freestanding graphene prepared by the platinum-metal catalysis method remains clean, even after re-exposure to ambient pressure and subsequent wet deposition of nanometre-sized gold rods. We present low-energy electron holograms of gold nanorods on graphene and cross-validate the presence of the nanorods by scanning electron imaging of the very same sample. Moreover, we compare the appearance of the rods when either imaged with low-energy electron holography or by means of a scanning electron microscope (SEM).

**Materials and Methods**

Ultraclean freestanding graphene, covering holes of 500nm in diameter milled in a silicon nitride membrane, is prepared by the platinum-metal catalysis method, described in detail recently elsewhere [18]. Thereafter, the cleanliness of the as-prepared graphene is inspected in a low-energy electron point source microscope operated under UHV conditions (Figure 1). In this holographic setup, inspired by the Gabor's original idea of inline holography [21–23], a sharp (111)-oriented tungsten tip acts as source of a divergent beam of highly coherent electrons [24–27]. The electron emitter can be brought as close as 200nm to the sample with the help of a 3-axis nanopositioner. Part of the electron wave impinging onto the sample is elastically scattered and represents the object wave, while the un-scattered part of the wave represents the so-called reference wave [28]. At a distant detector, the interference pattern between object wave and reference wave – the hologram – is recorded. The magnification in the image is given by the ratio of detector-tip-distance to sample-tip-distance and is typically of the order of $10^6$.

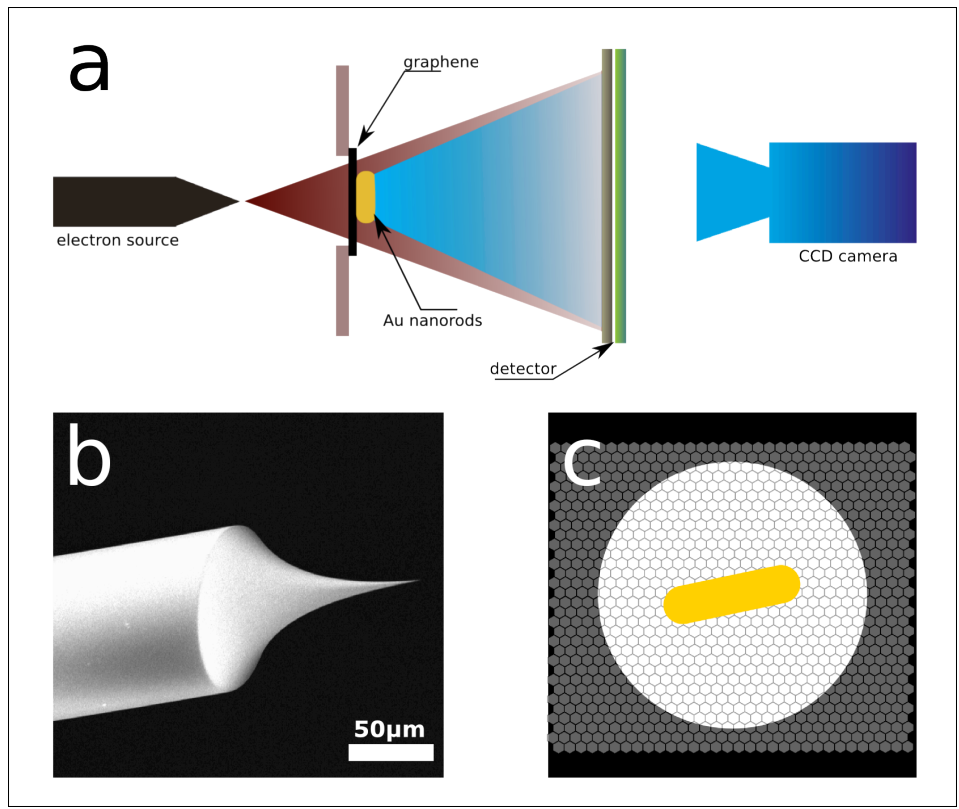

*Figure 1: **a.** Scheme of the experimental setup of low-energy electron holography. The source-sample distance amounts to typically 100-1000nm which leads to kinetic electron energies in the range of 50-250eV and the sample to detector distance is 68mm. The electron detector is 75mm in diameter large, which represents an acceptance angle of 29° **b.** SEM image of an electrochemically etched W(111) tip acting as field-emitter of a divergent coherent low-energy electron beam. **c.** Schematic illustration of the sample geometry with a gold nanorod lying on ultraclean freestanding graphene suspended over a round hole.*

**Results and Discussion**

Figure 2(a) shows an example of a hole of 500nm in diameter covered by a single layer of ultraclean graphene, imaged by low-energy electrons. Only the observation of interference fringes, arising due to the presence of a few hydrocarbons less than 1nm in size, reveals the existence of graphene covering the hole [18]. The cleanliness of the as-prepared graphene has also been investigated by means of high-resolution transmission electron microscopy (TEM) at 80kV in order to give the reader the possibility to compare the quality of the cleanliness with former TEM results. Figure 2(b) shows a TEM image of graphene, uniformly covering the entire freestanding region, and it is only by imaging the hexagonal atomic arrangement (Figure 2(c)) that the presence of graphene can reliably be confirmed.

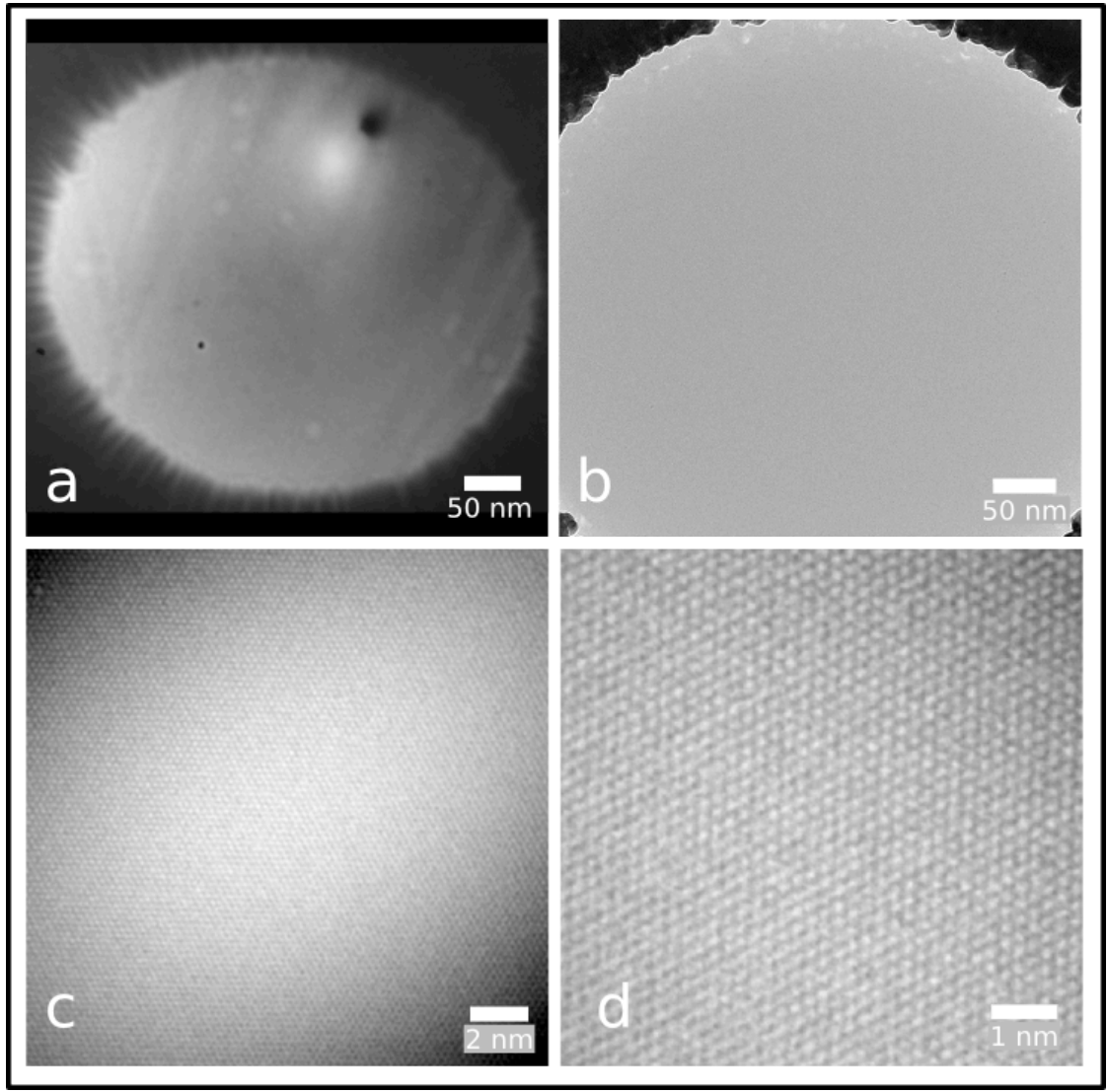

*Figure 2:* ***a.*** *Low-energy (62eV) electron transmission image of ultraclean freestanding graphene covering a hole of 500nm in diameter milled in a silicon nitride membrane.* ***b.*** *80kV TEM imaging of ultraclean graphene covering a hole of 500nm in diameter milled in a silicon nitride membrane.* ***c.*** *High-resolution TEM imaging of a 19x19nm$^2$ region of ultraclean freestanding graphene, the unit cell arrangement is visible and the atomic cleanliness of the graphene is conserved over the whole freestanding area. d. High-resolution TEM imaging of a 8x8nm$^2$ region of ultraclean freestanding graphene. TEM data by courtesy of Gerardo Algara-Siller from the University of Ulm.*

For the deposition of gold nanorods, a graphene sample prepared as described above is taken out of the low-energy electron microscope. Under ambient conditions, a drop of a 0.5nM gold nanorod aqueous solution [29] is subsequently applied onto the graphene (Figure 3(b)). A few seconds were given for the rods to sediment before the excess water was removed by using a filter paper (Figure 3(c)). Prior to the re-introduction of the sample into the electron microscope, the sample is kept at 200°C for 30min.

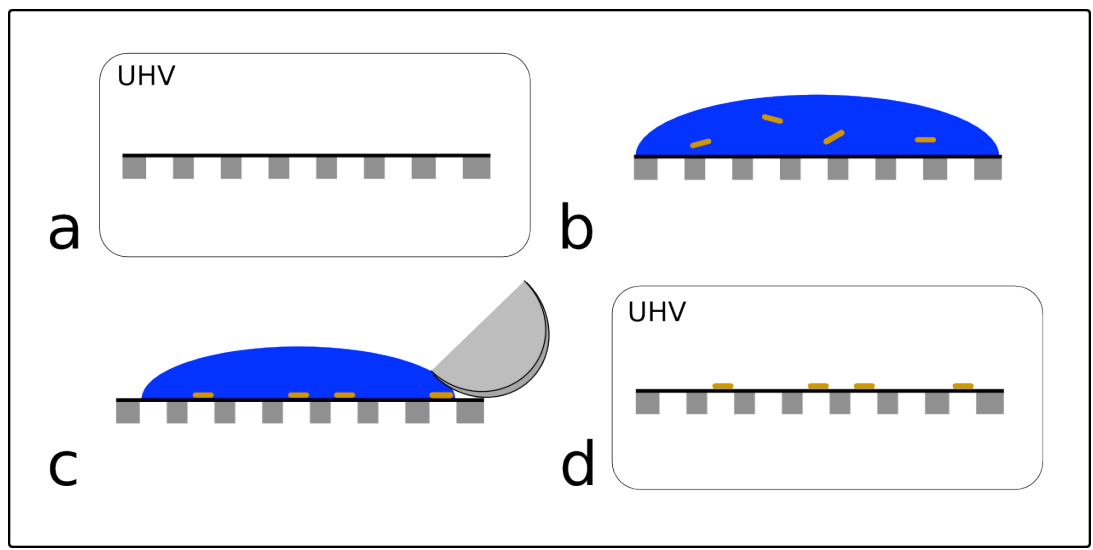

*Figure 3: **a.** The cleanliness of the graphene sample carrier is first inspected by means of low-energy electron holography under UHV conditions. **b.** A drop of the solution containing gold nanorods is applied onto the substrate. **c.** After waiting a few seconds for the sedimentation of the gold nanorods onto the graphene, the excess water is removed with a filter paper. **d.** Before the re-insertion in the UHV low-energy electron microscope, the sample is re-heated to 200°C for 30min. The nanorods are now ready to be imaged by means of low-energy electron holography.*

Figure 4(a) shows an electron hologram of gold nanorods on freestanding graphene recorded with 93eV kinetic energy electrons. The graphene surrounding the rods remained clean even after the re-exposure to ambient pressure and the deposition of the gold particles from the liquid phase. In Figure 4(b) a SEM image (7kV) of the very same sample is presented. The nanorods can be associated one-to-one with the objects observed in the holographic image presented in Figure 4(a). The yield of secondary electrons produced by the graphene substrate is so low that the rods, shown in Figure 4(b), seem to levitate, demonstrating the utility of graphene as a substrate for scanning electron microscopy. A high-magnification hologram (58eV) of a gold nanorod is presented in Figure 4(c) along with its reconstruction, see Figure 4(d), obtained as described in [30,31]. The object presented in these two images is the very same gold nanorod observed in Figure 4(a) in the low right corner. The remaining interference fringes that can be observed around the object in the reconstruction (Figure 4(d)) are due to the presence of the out-of-focus twin image [32]. The size of the rod in Figure 4(d) accounts for a width of 30nm and a length of 72nm. While the length in the holographic reconstruction image matches perfectly the length that can be measured in the SEM image, a discrepancy opens up when one compares the width measured in the two images (30nm in the holographic reconstruction and 21nm in the SEM image). We associate this discrepancy with the fact that the gold rods feature an organic coat in order to be soluble in aqueous solution [29]. This organic layer, however, is only present along the rods but not at the face sides. The several-nanometre thick methyl-shell cannot be imaged in an SEM because of the low contrast that it produces and because of the radiation damage provoked by the high-energy electrons. In low-energy electron holography the electron scattering cross-section depends only very weakly on the atomic number, therefore, the organic shell yields a substantial signal. Similarly, in Figure 4(a) a graphene nanoribbon, most likely produced during the CVD growth process, can be detected by low-energy electron holography while it cannot be seen in the SEM image.

In Figure 4(d), a plateau at the upper part of the rod and a tip-like shape at the lower part are visible. These observations are in accordance with TEM observations on similar gold nanorods [33]. The terminations of the rods can adopt different configurations in order to reduce the surface free energy. The plateau shape corresponds to a termination with a {001} facet. The tip shape termination is probably due to a very small {001} terrace surrounded by extended {111} facets forming the conical shape of this termination [33,34]. The geometrical details of the termination of the gold nanorods are accessible in low-energy electron holography imaging; the resolution of an SEM is insufficient to reveal them.

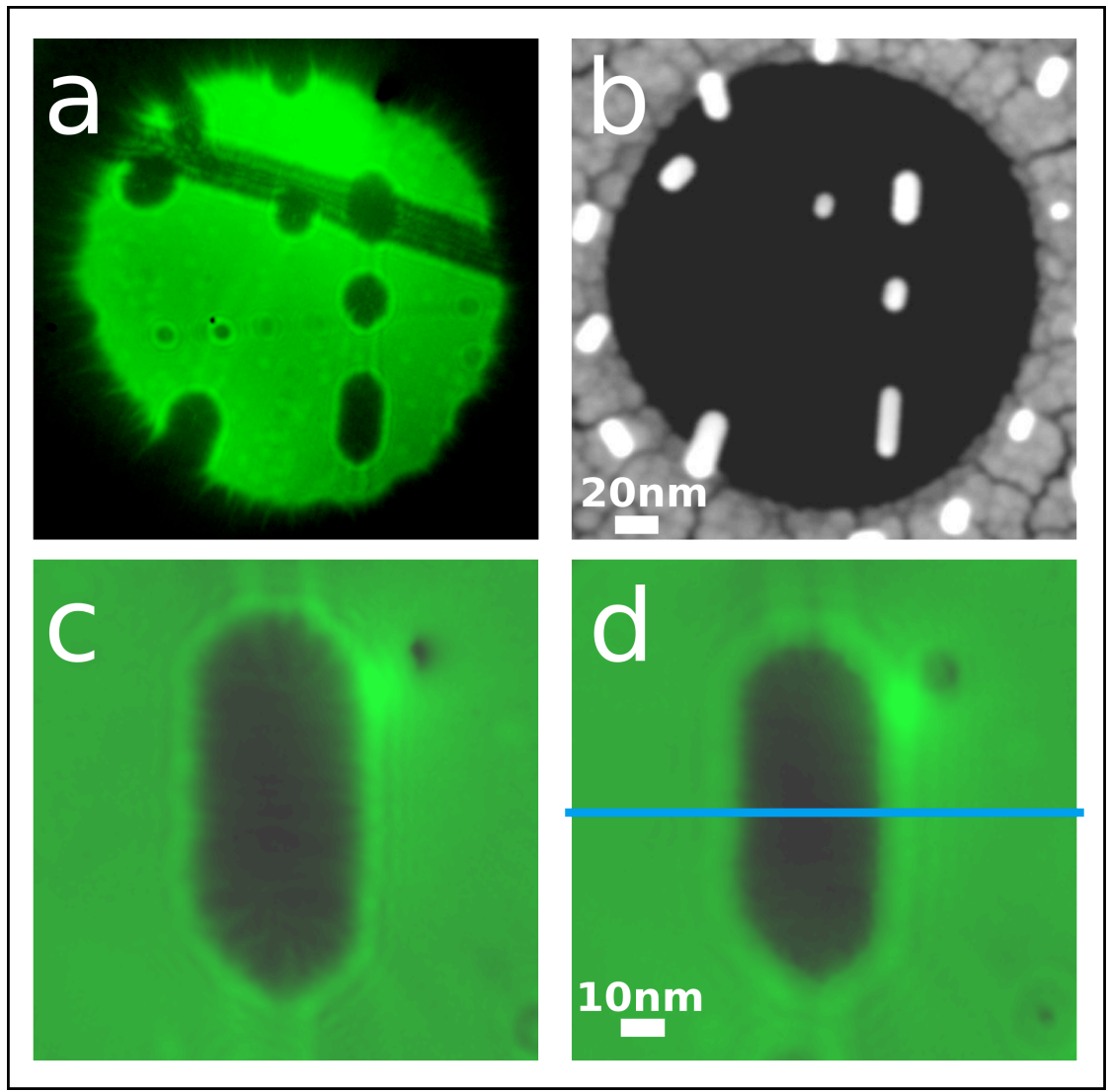

Figure 4: **a.** Low-energy electron hologram (93eV) of gold nanorods lying on freestanding graphene. The graphene remained clean even after the deposition of the nanorods. **b.** SEM image (7kV) of the very same sample presented in **a. c.** High-magnification low-energy electron hologram (58eV) of the nanorod on the lower right side presented in **a.** and **b. d.** The shape of the gold nanorod is reconstructed from the hologram in **c** at a source-sample distance of 182nm**.** An intensity profile along the blue line is displayed in Figure 5.

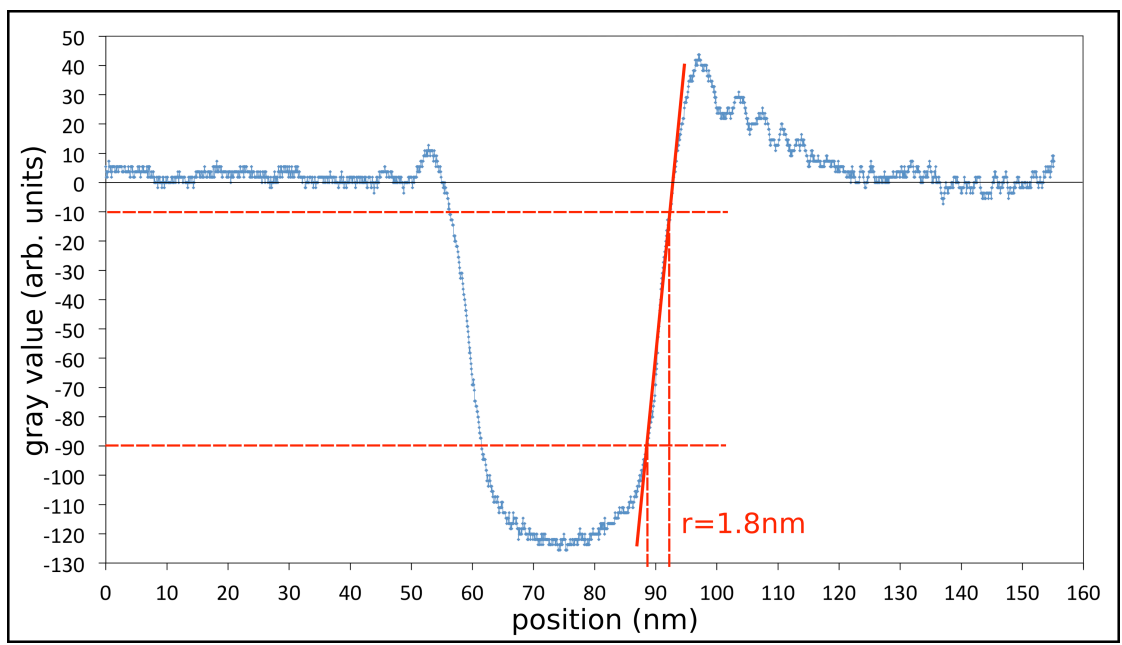

*Figure 5: Intensity profile across the whole reconstructed area of a gold nanorod lying on graphene (Figure **4d**, blue line). The resolution in the image is estimated by applying a linear fit to the edge response and amounts to 1.8nm.*

In Figure 5, an intensity profile along the blue line in Figure 4(d) is displayed. The intensity oscillations that can be observed around the region corresponding to the nanorod are due to presence of the out-of focus twin image.

In order to assign an upper limit for the resolution obtained in the reconstruction of the gold nanorod (Figure 4(d)), one may assume a sharp edge between the nanorod and the substrate. An estimate for the resolution is then given by the distance between 10 and 90% of the maximum intensity [35] measured at the edge and amounts to 1.8nm. This estimation is rather conservative because of the presence of the organic shell around the gold nanorod. Certainly, this shell does not represent an ideal sharp edge but the slight transmission through the organic shell tends to underestimate the resolution power. A resolution of 1–1.5 nanometres is probably a more realistic estimation for the resolution that can be obtained in our low-energy electron holography set-up.

**Conclusions**

In summary, we have shown that ultraclean graphene prepared by the platinum-metal catalysis method remains clean even after re-exposure to ambient pressure and wet deposition of gold nanorods. The outstanding cleanliness of the substrate allows gold nanorod imaging by means of low-energy electron holography with a resolution of 1–2nm. The organic shell surrounding the gold nanorods can be observed with low-energy electron holography while it is not detected in SEM imaging. Combining nanometre resolution, an enhanced scattering cross-section for low atomic number elements and the lack of radiation damage by low-energy electrons to biomolecules on a graphene sample carrier, defines low-energy electron holography as a powerful tool for structural biology at the single molecule level.

**Acknowledgements**

The authors would like to thank Gerardo Algara-Siller and Professor Ute Kaiser from the University of Ulm for the high-resolution TEM imaging of ultraclean graphene. The authors are also grateful for financial support by the Swiss National Science Foundation.